\documentclass[twocolumn, 10pt, pra, aps, showpacs, reprint, amsmath, amssymb, superscriptaddress, nofootinbib]{revtex4-1}

\usepackage{graphicx}

\begin{document}

\title{Random-laser dynamics with temporally modulated pump}

\author{Stefan Bittner}
\author{Sebastian Knitter}
\author{Seng Fatt Liew}
\author{Hui Cao}
\email{Corresponding author: hui.cao@yale.edu}
\affiliation{Department of Applied Physics, Yale University, New Haven, Connecticut 06520, USA}

\date{\today}

\begin{abstract}
We present an experimental study of the effects of temporal modulation of the pump intensity on a random laser. The nanosecond pump pulses exhibit rapid intensity fluctuations which differ from pulse to pulse. Specific temporal profiles of the pump pulses produce extraordinarily strong emission from the random laser. This process is deterministic and insensitive to the spatial configuration of the scatterers and spontaneous emission noise.
\end{abstract}


\maketitle

\section{Introduction}

In a random medium, disorder-induced light scattering can provide feedback for lasing action. The random laser is a complex nonlinear system involving many spatial modes which interact nonlinearly with the gain medium. The modal interactions can be tuned by shaping the pump beam spot \cite{Leonetti2011, Leonetti2012}. Typically a random laser emits light at multiple frequencies, but an adaptive shaping of the spatial pump profile enables singlemode lasing at a predetermined frequency \cite{Bachelard2012, Bachelard2014}. Spatial modulation of the pump intensity can make the random-laser emission directional \cite{Hisch2013} and also control the number of lasing modes \cite{Ge2015, Ge2017}. So far, temporal modulation of the pump has not been explored for random lasers. A technical challenge is the high speed required for temporal modulation. Prior experimental studies with pulsed excitations have revealed fast dynamic response of random lasers \cite{Lawandy1994, Sha1994, Siddique1996, Noginov1996, Soest2001, Soukoulis2002, Anglos2004, Sun2007, Molen2009, Iparraguirre2013, Gorbunov2015, Shi2017a, Kumar2017}, and the pump modulation needs to be on a comparable time scale. 

Here we present an experimental study of the effects of temporal modulation of the pump intensity on random lasers. The pump pulses exhibit rapid intensity fluctuations, and these fluctuations differ from pulse to pulse. Experimentally, we record the time traces of each pump pulse and the corresponding emission pulse. By measuring a large number of pulses, we investigate the dynamic response of random lasers to various pump modulations in time. Experimentally, we find that specific temporal profiles of the pump pulse induce extraordinarily strong emission from the random laser. This work paves the way for controlling random lasers by temporal modulation of the pump. 

\section{Random lasing experiment}

We fabricate ZnO nanoparticle films by spin coating on a glass substrate. The mean particle diameter is about $100$~nm, and the film thickness is approximately $10~\mu$m. Scanning electron microscope (SEM) images show random fluctuations of the film thickness and particle density on a length scale of $10$--$50~\mu$m across the film due to particle clustering during the fabrication process. This results in a variation of light transmission across the sample on the same length scale. The transport mean free path varies between $1$ and $2~\mu$m, which is much smaller than the film thickness. Thus light transport is in the diffusive regime, assuming that absorption is negligible. 

\begin{figure}[tb]
\includegraphics[width = 6.4 cm]{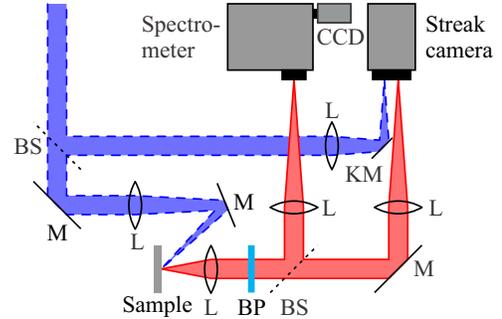}
\caption{(Color online) Sketch of the experimental setup (not to scale). The pump beam ($355$~nm) is depicted with blue dashed lines and the random-laser emission ($\approx 385$~nm) with red solid lines. M: mirror, KM: knife-edge mirror, L: lens, BS: pellicle beam splitter, BP: bandpass filter at $385$~nm, CCD: CCD camera.}
\label{fig:setup}
\end{figure}

To obtain lasing at room temperature, we optically pump the ZnO film with the third harmonic ($\lambda = 355$~nm) of a Q-switched Nd:YAG laser (Continuum Minilite TEM00) at a repetition rate of $10$~Hz. The pulse width is approximately $5$~ns, and the diameter of the pump spot on the sample surface is about $400~\mu$m. A schematic drawing of the experimental setup is shown in Fig.~\ref{fig:setup}. The emission from the ZnO film is collected by a microscope objective and any scattered pump light removed by a bandpass filter. The collected emission is split and focused onto the entrance slits of an imaging spectrometer (Acton Research SP300i) equipped with a CCD camera (Andor Newton DU920P-BEX2-DD) and a streak camera (Hamamatsu C5680) with a fast sweep unit (M5676), respectively. In addition, a sample of the pump beam is split off with a beam splitter and fed to the streak camera in parallel with the laser emission. Thus, the spatiospectral images of the emission are measured in parallel with the temporal dynamics of the pump and the random-laser emission. All data presented in the following are single-shot measurements. 

\begin{figure}[tb]
\includegraphics[width = 8.4 cm]{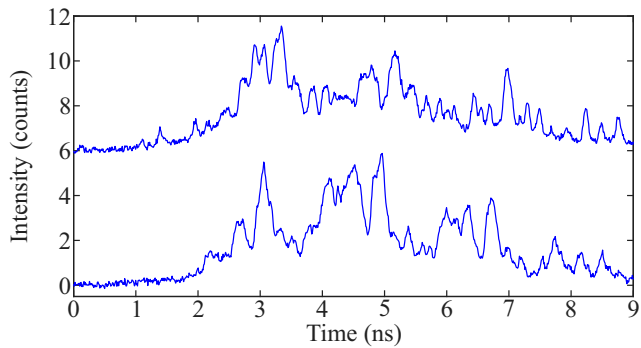}
\caption{Pump	pulse shape. Two examples of pump pulses showing very different temporal fluctuations. The two curves are offset vertically by six counts.}
\label{fig:PumpPulseExmpls}
\end{figure}

Before measuring the ZnO emission, we first characterized the pump pulses with the streak camera. Figure~\ref{fig:PumpPulseExmpls} shows the temporal traces for two pump pulses. Both are highly structured with intensity fluctuations on the order of a few hundred picoseconds. The strong modulations are caused by the interference of multiple longitudinal modes of the pump laser. The temporal autocorrelation function of the pump pulses shows that the average spacing of adjacent intensity peaks in a single pulse is $1.8$~ns, corresponding to the round-trip time of the pump laser cavity. Each pump pulse has a distinct temporal profile, since the relative phases of the longitudinal modes change randomly from shot to shot. 

\section{Extraordinarily strong emission pulses}

\begin{figure}[tb]
\includegraphics[width = 8.4 cm]{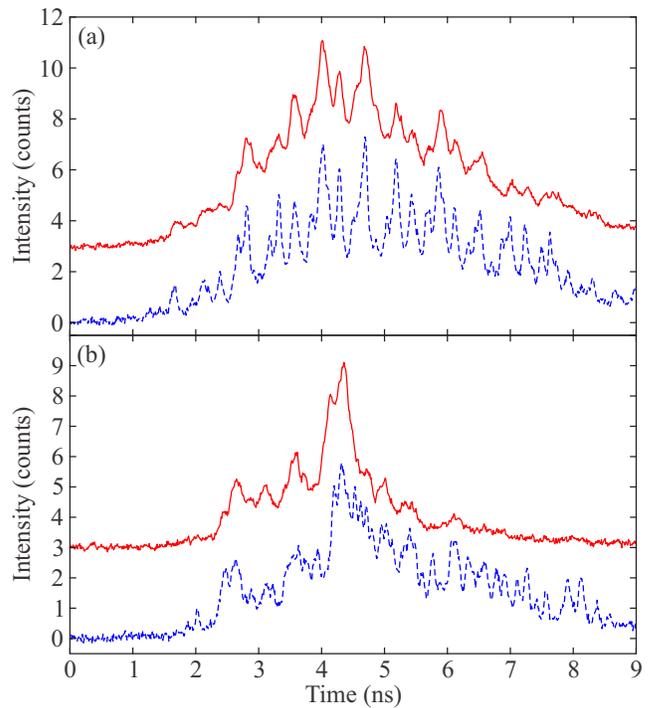}
\caption{(Color online) Emission pulse shape. (a)~For pump pulse energy $U_p = 50~\mu$J (below lasing threshold), the temporal fluctuations of the emission intensity (red solid line) follow that of the pump intensity (blue dashed line). (b)~For pump pulse energy $U_p = 170~\mu$J (above lasing threshold), the emission pulse is cut short by quenching during the later part of the pump pulse. The curves for the emission pulses are vertically offset by three counts for clarity.}
\label{fig:EmissionPulseExmpls}
\end{figure}

Next we perform temporal measurements of the emission from the ZnO film. Since the temporal profile of the pump pulse changes from one shot to the next, we simultaneously record the pump and the emission traces with the streak camera (see Fig.~\ref{fig:setup}). Figure~\ref{fig:EmissionPulseExmpls}(a) shows the time traces of emission (red solid line) and pump (blue dashed line) for the same shot at a pump pulse energy of $U_p = 50~\mu$J. The emission intensity varies in time and follows the modulation of the pump intensity. The emission intensity time trace is identical for different positions on the sample surface. The emission pulse has the same length as the pump pulse. However, when the pump pulse energy is increased to $U_p = 170~\mu$J, the emission pulse is shorter than the pump pulse [see Fig.~\ref{fig:EmissionPulseExmpls}(b)]. This shortening is also observed for a monocrystalline bulk ZnO sample and is attributed to the quenching of the ZnO emission. During the experiments, we monitor the speckle pattern of the pump light scattered from the nanoparticle film. The speckle pattern remains unchanged with increasing pump power until the sample is damaged by the pump laser. We keep the pump level below the damage threshold, and the ZnO emission is fully recovered when the next pump pulse arrives $100$~ms later. 

\begin{figure}[tb]
\includegraphics[width = 8.4 cm]{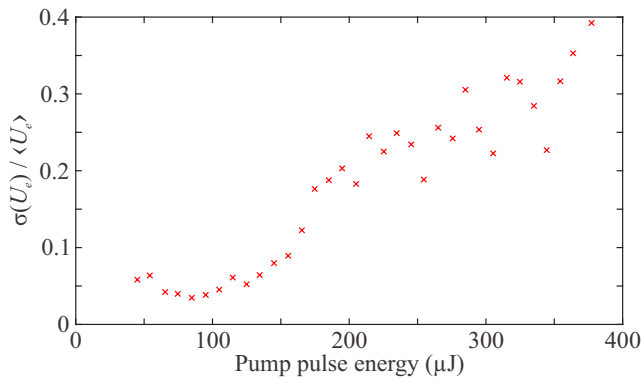}
\caption{Shot-to-shot fluctuations of emission pulse energy. The standard deviation of the emission pulse energy, normalized by the mean, $\sigma(U_e)/\langle U_e \rangle$, is plotted as a function of the pump pulse energy. The rapid increase above the lasing threshold $U_{p, th} \simeq 130~\mu$J indicates strong fluctuations of the lasing emission.}
\label{fig:LIcurveVar}
\end{figure}

In spite of the emission quenching during the later part of the pump pulses, we observe lasing during the earlier part when the pump pulse energy exceeds the threshold of $U_{p, th} \simeq 130~\mu$J. Discrete peaks appear in the emission spectrum, and the emission pulse energy $U_e$ fluctuates strongly from shot to shot. Figure~\ref{fig:LIcurveVar} shows the standard deviation of the emission energy $\sigma(U_e)$, normalized by the mean $\langle U_e \rangle$, over many pulses at the same pump level. The ratio $\sigma(U_e) / \langle U_e \rangle$ remains around $0.05$ below the lasing threshold, but it increases rapidly above the lasing threshold. 

\begin{figure}[tb]
\includegraphics[width = 8.4 cm]{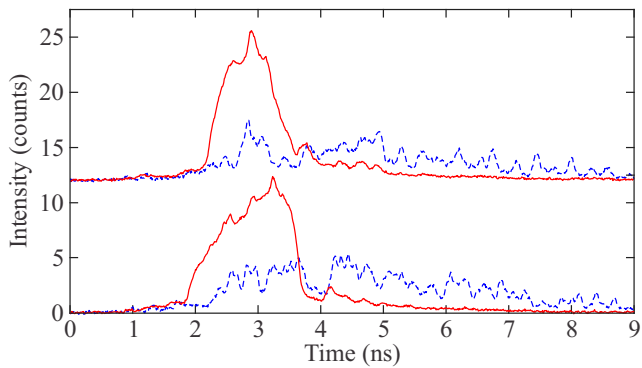}
\caption{(Color online) Extraordinarily strong emission pulses. Instead of following the temporal modulation of the pump intensity (blue dashed lines), the emission intensity (red solid lines) rises sharply at the beginning of the pump pulse. The pump pulse energy is $U_p = 240~\mu$J. The two examples are offset vertically by $12$~counts.}
\label{fig:StrongPulseExmpls}
\end{figure}

The strong fluctuations of the emission pulse energy originate from emission pulses with extraordinarily high intensities. Figure~\ref{fig:StrongPulseExmpls} presents two examples. In contrast to the regular emission pulses, an extraordinarily intense emission pulse does not follow the temporal modulation of the pump intensity; instead, its intensity rises sharply at the beginning of the pump pulse. 

\begin{figure}[tb]
\includegraphics[width= 8.4 cm]{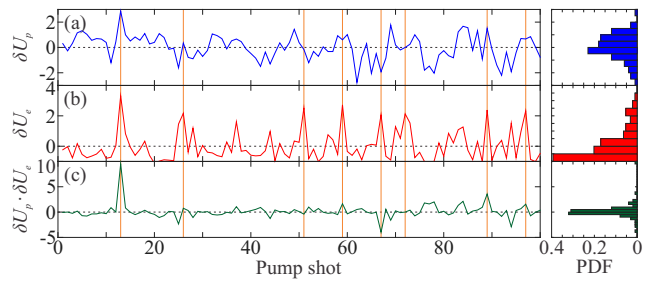}
\caption{Correlation between pump and emission energy fluctuations, $\delta U_{p,e}$, recorded simultaneously for a sequence of $100$~shots. The right panel shows the corresponding probability density function (PDF) of $\delta U_{p,e}$. (a, b)~Emission pulse energy fluctuations in (b) are mostly uncorrelated with the pump pulse energy fluctuations in (a). The PDF for $\delta U_{e}$ in (b) shows a long tail of the emission intensity distribution, in contrast to the nearly symmetric PDF for $\delta U_{p}$ in (a). (c)~The product $\delta U_{p} \delta U_{e}$ is close to zero for most pulses, confirming the lack of correlation between the pump and emission fluctuations. Strong emission pulses are marked by vertical orange lines. The horizontal dotted black lines indicate $0$.}
\label{fig:emission-fluctuations}
\end{figure}

After collecting data over many pulses, we find that about $5$--$10\%$ of the pump pulses produce emission pulses with an energy greater than the mean by $2 \sigma$, i.e., $U_e > \langle U_e \rangle + 2 \sigma(U_e)$. Their occurrence is not caused by the fluctuations of the pump pulse energy as shown in Fig.~\ref{fig:emission-fluctuations}, where the fluctuations $\delta U_{p,e} = [U_{p,e} - \langle U_{p,e} \rangle] / \sigma(U_{p,e})$ of pump and emission pulse energy [Figs.~\ref{fig:emission-fluctuations}(a) and \ref{fig:emission-fluctuations}(b), respectively] for a sequence of $100$ shots are presented. The product $\delta U_p \delta U_e$ of pump and emission fluctuations in Fig.~\ref{fig:emission-fluctuations}(c) vanishes on average, indicating that the shot-to-shot fluctuations of the emission pulse energy are uncorrelated with the fluctuations of the pump pulse energy. Some of the intense emission pulses are even produced by pump pulses with an energy below the average (marked by the vertical orange lines in Fig.~\ref{fig:emission-fluctuations}). 

\section{Spatiospectral measurements of random lasing emission}

A further characterization of the lasing emission is conducted with spatiospectral measurements. The surface of the ZnO film is imaged onto the entrance slit of the imaging monochromator (see Fig.~\ref{fig:setup}). The entrance slit samples the emission from a narrow stripe within the pump area, and the emission is spectrally dispersed by the monochromator. The two-dimensional (2D) spatiospectral image, recorded by the CCD camera, shows the spatially resolved emission spectrum. 

\begin{figure*}[tb]
\includegraphics[width = 14 cm]{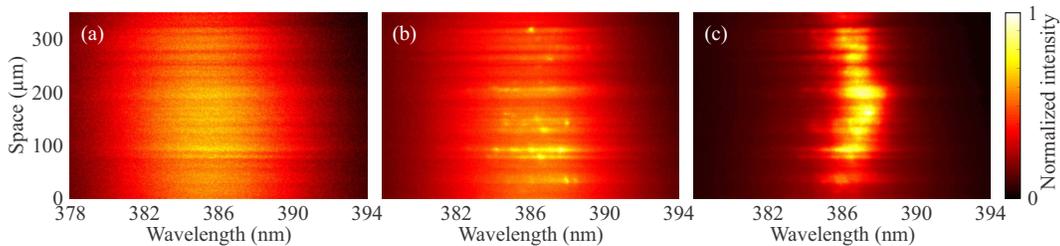}
\caption{(Color online) Spatiospectral measurements of the random-laser emission. (a)~2D spatiospectral image of the ZnO emission at a pump pulse energy of $U_p$ = $50~\mu$J. The spontaneous emission is stronger from the denser and/or thicker regions of the ZnO film. (b)~Spatiospectral image of ZnO emission at $U_p$ = $220~\mu$J. Lasing occurs in some of the denser and/or thicker regions at distinct wavelengths. (c)~Spatiospectral image of an extraordinarily strong emission pulse at $U_p$ = $310~\mu$J.}
\label{fig:spatio-spectral}
\end{figure*}

At low pumping level, the spontaneous emission is stronger from the denser and/or thicker regions of the ZnO film [see Fig.~\ref{fig:spatio-spectral}(a)]. Above the lasing threshold, the spatiospectral image in Fig.~\ref{fig:spatio-spectral}(b) shows isolated regions of lasing emission from the sample. This result indicates that lasing occurs in some of the denser and/or thicker regions, and the lasing wavelengths vary from one region to another. For the extraordinarily intense emission pulses, most denser and/or thicker regions within the pump beam spot exhibit intense lasing emission, as shown in Fig.~\ref{fig:spatio-spectral}(c). The emission spectrum is smooth and overlaps for different regions, although the spectral width varies from one region to the other. 

\begin{figure}[tb]
\includegraphics[width = 8.4 cm]{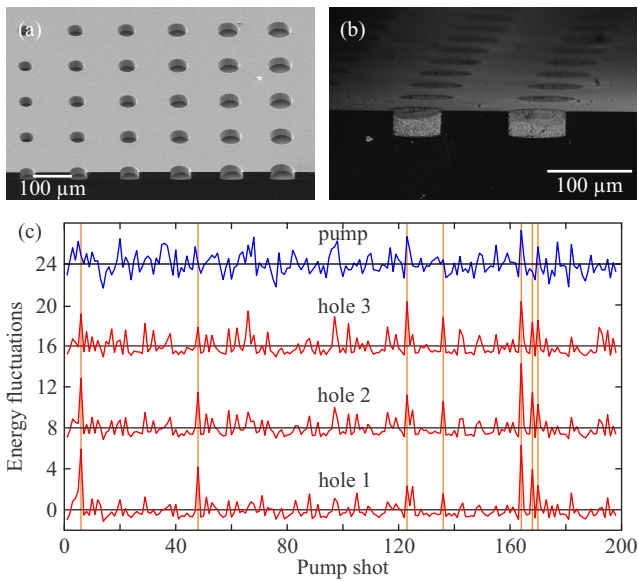}
\caption{(Color online) Random lasing in isolated holes. (a)~SEM image of an array of empty holes etched into silicon. (b)~SEM images of holes filled with ZnO nanoparticles. (c)~Energy fluctuations $\delta U_{p,e} = [U_{p,e} - \langle U_{p,e} \rangle] / \sigma(U_{p,e})$ of pump pulses (top blue line) and emission pulses (red lower lines) for three different holes in the sample (labeled holes \ $1$--$3$). Strong emission pulses (marked by the orange vertical lines) appear simultaneously in all holes. The different curves are vertically offset by $8$, and the horizontal black lines indicate zero for each curve.}
\label{fig:SeparatedEmitters}
\end{figure}

The fact that the strong emission occurs simultaneously in many locations of the sample suggests the intense lasing emission might result from a collective effect, caused by seeding of the emission from one denser and/or thicker region to the neighboring ones \cite{Leonetti2013b, Leonetti2013a}. To test this possibility, we fabricate another sample shown in Figs.~\ref{fig:SeparatedEmitters}(a) and \ref{fig:SeparatedEmitters}(b). An array of holes (diameter $50~\mu$m, depth $28~\mu$m, spacing $130~\mu$m) is etched in a silicon wafer [Fig. \ref{fig:SeparatedEmitters}(a)]. Each hole is filled with ZnO nanoparticles [Fig.~\ref{fig:SeparatedEmitters}(b)]. The strong absorption of ZnO emission by the silicon prohibits seeding of the emission from the ZnO in one hole to the neighboring one. 

When pumping multiple holes at the same time, extraordinarily strong emission pulses appear in all the holes simultaneously. Figure~\ref{fig:SeparatedEmitters}(c) shows the emission pulse energy fluctuations from three adjacent holes for successive pump pulses: when a strong emission event is observed in one hole, it is also registered in the other two holes, as highlighted by the orange vertical lines. Since any interaction between the lasing modes in the different holes is excluded, a collective effect is ruled out as the cause of extraordinarily strong emission pulses. 

Another possible origin of the extraordinarily strong emission events are the fluctuations of the initial spontaneous emission that induces lasing. However, the spontaneous emission of ZnO films in different holes should be uncorrelated. It is thus very unlikely that the initial spontaneous emission is unusually strong in all holes for a single shot to induce intense lasing emission simultaneously. Unusually strong emission events from random lasers have also been attributed to so-called "lucky photons" that are strongly amplified due to a long path length in the disordered medium, which can result in L\'evy-distributed intensity statistics \cite{Lepri2007, Uppu2012, Uppu2013, Ignesti2013}. This mechanism, however, is excluded here, since it cannot explain the correlated appearance of strong emission from different holes and across the whole spectrum. 

\section{Temporal modulation of the pump intensity}

The remaining apparent cause for the strong emission fluctuations is the temporal modulation of the pump intensity. We hence measure the emission fluctuations for a nanosecond pump laser featuring a smooth temporal profile that is stable from shot to shot (see Appendix~\ref{sec:smoothPump}). In this case, the emission pulse energy has little fluctuation from pulse to pulse as shown in Fig.~\ref{fig:LIcurveSeeded}. 

Therefore, in order to check whether specific temporal profiles of the pump pulse can generate extraordinarily strong emission pulses, we split each pump pulse from the Minilite nanosecond laser with a beam splitter and delay one copy of the pump pulse with respect to the other by about $13$~ns. The two pulses are focused on two different holes of the silicon wafer shown in Fig.~\ref{fig:SeparatedEmitters} so they excite ZnO nanoparticles in different holes at different times. In other words, the two pump pulses, which have identical temporal profiles, have neither spatial nor temporal overlap on the ZnO sample. 

\begin{figure}[tb]
\includegraphics[width = 8.4 cm]{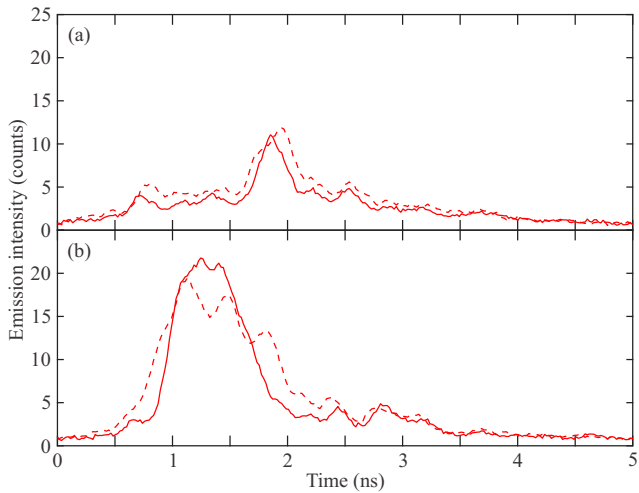}
\caption{Direct comparison of the emission time traces from ZnO nanoparticles in two different holes. They are pumped by identical pulses which are delayed by $13$~ns (the time delay is subtracted in the plot). The first (solid) and the second (dashed) emission pulses have a nearly identical temporal profile, for both regular (a) and extraordinarily strong (b) emission events.}
\label{fig:delayPulseTimetraces}
\end{figure}

We measure the emission from the two holes to investigate potential correlations. Figure~\ref{fig:delayPulseTimetraces} shows two examples of the first and second emission pulses (where the time delay is subtracted). The first example in Fig.~\ref{fig:delayPulseTimetraces}(a) is a regular event, and the second example in Fig.~\ref{fig:delayPulseTimetraces}(b) is a strong event. In both cases, the first and second pulses show good agreement and exhibit very similar temporal profiles. 

For a quantitative analysis, we collect the statistics of the emission pulse energies from the two holes. Over the course of $500$ pump shots, $21$ ($27$) emission pulses from the first (second) hole have an energy exceeding the mean energy by $2 \sigma$. Had these strong emission events occurred independently of each other, only about $1$ out of $500$ shots would be expected to have strong emission from both holes. However, we observe strong emission from both holes simultaneously for $8$ out of $500$ shots, indicating the emission fluctuations from the isolated holes are highly correlated. The only source of this correlation is the common temporal profile of the pump pulse. 

\section{Discussion and Conclusion}

Although strong fluctuations of random-laser emission with nanosecond pulse pumping have been reported before \cite{Markushev2007a, Fallert2009, Zhu2012}, the underlying mechanism is not fully understood. We experimentally compare random lasing in the same ZnO nanoparticle films with both picosecond pumping (not shown) and nanosecond pumping. Although lasing does not reach the steady state with $30$-ps-long pump pulses, the lasing spectrum and emission pulse energy are much more stable and reproducible than those with $5$-ns-long pump pulses, in agreement with previously published data \cite{Cao2003a, Wu2004, Anglos2004}. This surprising result is attributed to large temporal fluctuations of the pump intensity during the $5$-ns pulses. Such fluctuations vary from pulse to pulse, and special temporal profiles of the pump intensity can produce extraordinarily strong lasing emission. This process is predominantly deterministic and insensitive to the spatial configuration of the scatterers or the initial spontaneous emission. 

It is known for conventional multimode lasers that temporal modulations of gain or loss can induce complex dynamic responses, including extreme events and crises \cite{Ogawa1987a, Otsuka1991, Mandel1993, Granese2016}. A random laser supports many lasing modes and is expected to display diverse behaviors for temporal modulation of the pump. Even with constant pumping, coherent instabilities are predicted for random lasers \cite{Andreasen2011}, and pronounced fluctuations, including the generation of extreme events, are observed in fiber lasers with randomly distributed feedback \cite{Gorbunov2015}. In our case, the pump pulses with varying temporal shape can drive the random laser to distinct trajectories in the high-dimensional phase space \cite{Perrone2014}, leading to strong fluctuations of the emission pulses. However, an analysis of the temporal profiles of the pump pulses could not identify any special feature of the pump pulses that produce intense emission. This may be due to the highly complex structure of the phase space of a multimode random laser. 

Unlike the periodic modulation commonly used for conventional lasers, the pump pulses in our experiment contain multiple driving frequencies. It has been shown for a singlemode laser that adding a second modulation frequency with small modulation amplitude can either reduce or enhance crisis-induced intermittency, depending on the phase difference between the two driving components \cite{Zambrano2006}. The multimode random laser has many more degrees of freedom than a singlemode laser; thus a slight modification of the temporal pump pulse profile, which might be hard to detect, could cause a dramatically different response. Detailed theoretical modeling of the random-laser dynamics with a modulated pump is needed to explain the experimental results. 

In conclusion, we show experimentally that temporal modulations of the pump intensity result in strong fluctuations of random-laser emission. Special temporal profiles of the pump pulse can generate extraordinarily strong emission. This result illustrates the potential of pump pulse shaping for significant enhancement of the output energy from a random laser and the possibility of creating giant emission pulses on demand. 

\appendix

\section{Nanosecond pumping with a single longitudinal-mode laser} \label{sec:smoothPump}

\begin{figure}[tb]
\includegraphics[width = 8.4 cm]{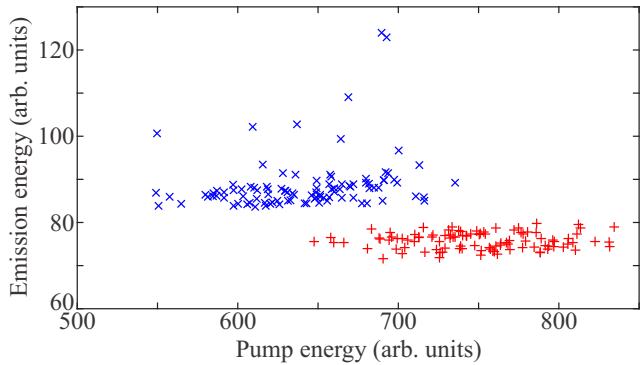}
\caption{Effect of pump pulse shape on random lasing. The emission pulse energy vs the pump pulse energy for smooth (red $+$) and temporally structured (blue $\times$) pump pulses. In the case of smooth pump pulses, the emission is weaker and has much less fluctuation.}
\label{fig:LIcurveSeeded}
\end{figure}

In a further experiment, a single longitudinal-mode pump laser was used in order to investigate the dynamics of the ZnO random lasers for nanosecond pulses that have a smooth temporal profile and are stable from shot to shot. We used a Surelite EX laser from Continuum that was seeded for single longitudinal-mode operation. The pulses were smooth and repeatable with a length of about $5$~ns for seeded operation, but acquired a jagged and unstable temporal profile with about $7$~ns length when the seed laser was deactivated, similar to the Minilite nanosecond laser. 

We record the spatiospectral images of the emission pulses with the spectrometer. The emission spectrum is smooth without discrete peaks, and the emission energy is stable when the pump laser is seeded. This is in sharp contrast to the random lasing behavior when the pump laser is unseeded. Figure~\ref{fig:LIcurveSeeded} displays the emission pulse energy versus pump pulse energy for both seeded and unseeded operation of the pump laser. The emission pulses are stronger and feature larger energy fluctuations for unseeded operation of the pump laser compared to the seeded operation with the same pump pulse energy. This result shows that the temporal profile of the pump pulse has a strong impact on random lasing in the ZnO nanoparticle films (cf.\ Ref.~\cite{Anglos2004}), and that hence the temporal profile of the pump pulse has a strong influence on the efficiency of converting the pump to light emission. 

\vspace{5 mm}

\begin{acknowledgments}
We thank Christian Vanneste and Jonathan Andreasen for extensive discussions and careful reading of the manuscript, as well as Kyungduk Kim, Hakan T\"ureci, Omer Malik, Patrick Sebbah, Yaron Bromberg, Brandon Redding and Cristina Masoller for useful discussions. S.K.\ thanks Yvette Just from Continuum for help with laser test measurements. This work is supported partly by the Office of Naval Research (ONR) with MURI Grant No.\ N00014-13-1-0649, and the Air Force Office of Scientific Research (AFOSR) under Grant No. FA9550-16-1-0416.
\end{acknowledgments}

\end{document}